\documentclass[prl,twocolumn,showpacs,floatfix,preprintnumbers,amsmath,amssymb,superscriptaddress]{revtex4}

\usepackage{graphicx}
\usepackage{dcolumn}
\usepackage{bm}
\usepackage{color}
\usepackage{color}
\usepackage[latin1]{inputenc}
\usepackage[urlcolor=blue]{hyperref}
\hypersetup{backref, colorlinks=true, linkcolor=blue, citecolor=blue}

\begin{document}

\title{High-Pressure Phase Diagram of NdFeAsO$_{0.9}$F$_{0.1}$: Disappearance of superconductivity on the verge of ferromagnetism from Nd moments} 

\author{Mahmoud Abdel-Hafiez}
\email{mahmoudhafiez@gmail.com}
\affiliation{Center for High Pressure Science and Technology Advanced Research, Beijing, 100094, China}


\author{M. Mito}
\affiliation{Graduate School of Engneering, Kyushu Institute of Technology, Fukuoka 804-8550, Japan}

\author{K. Shibayama}
\affiliation{Graduate School of Engneering, Kyushu Institute of Technology, Fukuoka 804-8550, Japan}

\author{S. Takagi}
\affiliation{Graduate School of Engneering, Kyushu Institute of Technology, Fukuoka 804-8550, Japan}

\author{M. Ishizuka}
\affiliation{Renovation Center of Instruments for Science Education and Technology, Osaka University, Osaka 560-0043, Japan}

\author{A. N. Vasiliev}
\affiliation{National University of Science and Technology (MISiS), Moscow 119049, Russia}
\affiliation{Moscow State University, Moscow 119991, Russia}
\affiliation{National Research South Ural State University, Chelyabinsk 454080, Russia}

\author{C. Krellner}
\affiliation{Institute of Physics, Goethe University Frankfurt, 60438 Frankfurt/M, Germany}

\author{H. K. Mao}
\affiliation{Center for High Pressure Science and Technology Advanced Research, Beijing, 100094, China}

\date{\today}

\begin{abstract}
We investigated transport and magnetic properties of NdFeAsO$_{0.9}$F$_{0.1}$ single crystal under hydrostatic pressures up to 50\,GPa. The ambient pressure superconductivity at $T_{c} \sim$ 45.4\,K is fully suppressed at $P_{c} \sim$ 21 GPa. Upon further increase of the pressure, the ferromagnetism associated with the order of rare-earth subsystem is induced at the border of superconductivity. Our finding is supported by the hysteresis in the magnetization $M$($H$) loops and the strong increase in the field cooled data, $M$($T$), toward low temperatures. We also show that the temperature evolution of the electrical resistivity as a function of pressure is consistent with a crossover from a Fermi-liquid to non-Fermi-liquid to Fermi-liquid. These results give access to the high-pressure side of the superconducting phase diagram in 1111 type of materials.
\end{abstract}


\maketitle


A key topic in current research of strongly correlated heavy fermion systems, high-$T_{c}$ cuprates and ferrates is the coexistence and competition between superconductivity and various electronic orders\cite{1,2}. In conventional superconductors, the electron-phonon interaction gives rise to the attraction between electrons with opposite momenta and opposite spins. This causes the superconductivity characterized by spin-singlet s-wave Cooper pairing and conservation of the time-reversal symmetry\cite{SC}. In contrast, the ferromagnetism breaks the time reversal symmetry, which makes these two phenomena antagonistic to each other. There are systems, however, where both superconductivity and ferromagnetism stems from the same electrons, although magnetism is to be suppressed prior to emergent superconducting (SC) order. In heavy fermion compounds UGe$_{2}$\cite{F2} and URhGe\cite{F3,F4} the superconductivity on the verge of ferromagnetism is understood in terms of magnetic interactions which presumes the spin-triplet pairing to be advantageous to the spin-singlet pairing. The antiferromagnetism is not excluded by superconductivity since the average values of magnetic induction and exchange field are negligibly small on the scale of the SC correlation length\cite{LN}. A unique coexistence of superconductivity, antiferromagnetism and ferromagnetism was observed in RuSr$_{2}$GdCu$_{2}$O$_{8}$, where all these phenomena were attributed to spatially separated CuO$_{2}$ planes, Gd and Ru magnetic moments\cite{F1}. Similarly, the coexistence of spatially separated ferromagnetism and superconductivity was observed in the ferroarsenite family\cite{F1,F5,F6}. Of utmost interest are the systems which evidence not just coexistence but interplay of these quantum cooperative phenomena in a single material.


The normal state of high-$T_{c}$ superconductors is quite unusual. The electrical resistivity vary with temperature in a peculiar way which deviates significantly from $\sim T^{2}$ dependence expected from Fermi-liquid (FL) theory of metals\cite{7,8,9,10}. Since non-Fermi-liquid (NFL) behavior is seen often above a SC dome, there is a consensus that its origin may hold the key to understand of the pairing mechanism in high-$T_{c}$ superconductors [15]. The studies on high-$T_{c}$ cuprates\cite{F1,8,9}, heavy fermion metals\cite{10}, organic Bechgaard salts\cite{N4}, and iron-based superconductors\cite{11,N5,N6,N7} imply that NFL behavior and high-$T_{c}$ SC dome favor proximity to magnetic order. This fact has led to proposals ascribing both NFL behavior and high-$T_{c}$ superconductivity to spin fluctuations close to a magnetic quantum critical point\cite{N8,N9}. At present, the microscopic mechanism of the NFL behavior and its relationship to high-$T_{c}$ superconductivity are still a matter of considerable debate.


\begin{figure*}[tbp]
\includegraphics[width=39pc,clip]{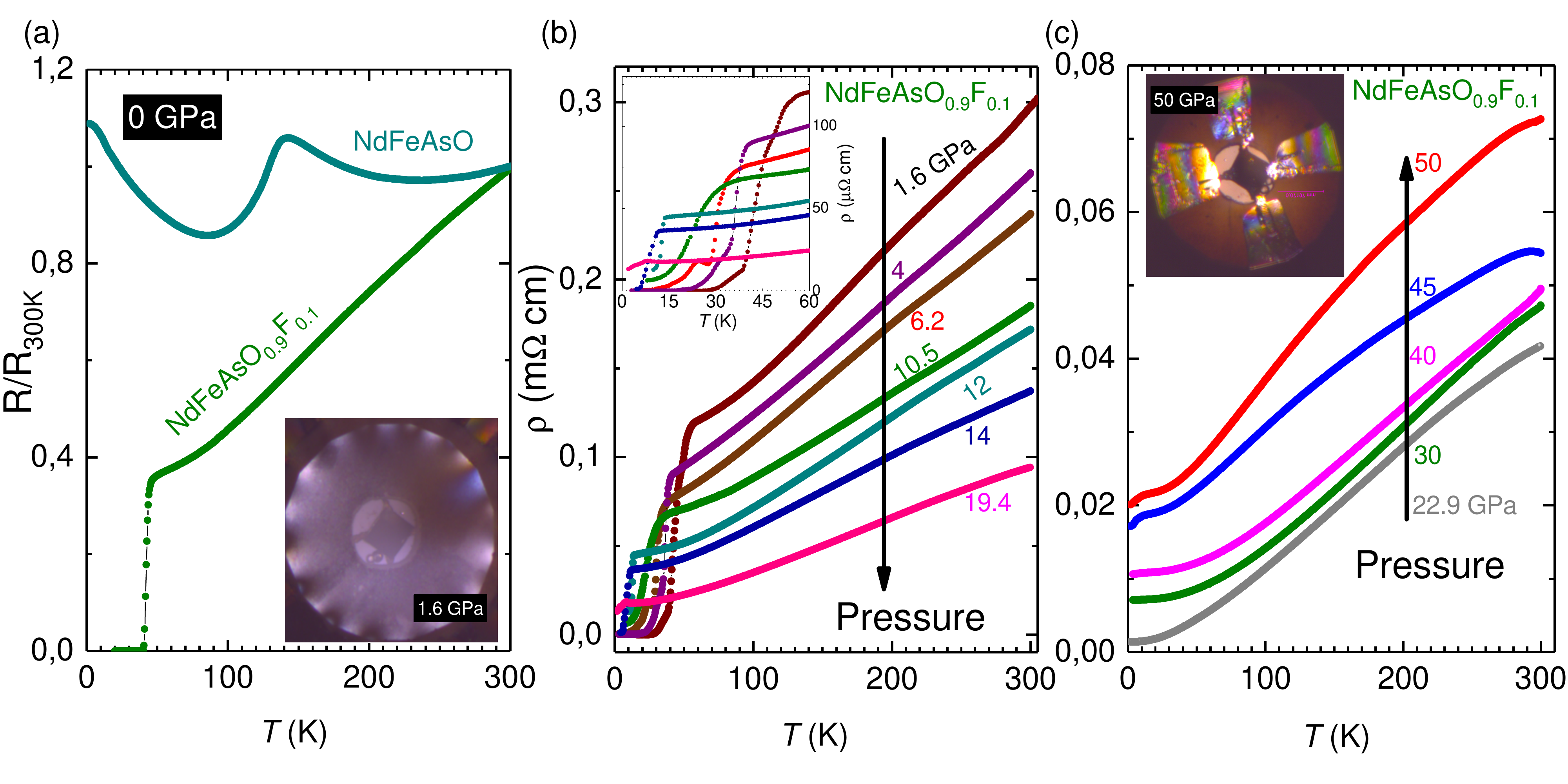}
\caption{\label{fig:wide}  \textbf{a}, illustrates the $T$-dependence of the in-plane resistance measurements upon heating of NdFeAsO$_{1-x}$F$_{x}$ single crystals at $x$ = 0 and 0.1. \textbf{b}, In-plane electrical resistivity $\rho$ versus temperature between 1.6 and 19.4\,GPa. The inset is the close-up of the low-temperature region, highlighting the SC transition. \textbf{c}, $T$-dependence of the resistivity curves at pressures between 23 and 50\,GPa. The insets of (\textbf{a}) and (\textbf{c}) present the image of NdFeAsO$_{0.9}$F$_{0.1}$ sample mounted in a diamond anvil cell at 1.6 and 50\,GPa, respectively.}
\end{figure*}

The 1111-type iron-based superconductors $Ln$FeAsO$_{1-x}$F$_{x}$ ($Ln$ stands for lanthanide) was the first material with a $T_{c}$ above 50\,K, other than cuprate superconductors\cite{3}. Change of rare earth ion or application of high pressure to the 1111-type materials\cite{4,5,6,5_1} has revealed a substantial influence of interionic distances on $T_{c}$. Thus, finding a way to tune the physical properties of NdFeAsO$_{0.9}$F$_{0.1}$ is important not only for understanding all of these interesting features but also for exploring the underlying mechanism of superconductivity in iron-based superconductors. Hydrostatic pressure is a widely used tool to study materials without changing their stoichiometry\cite{P1,P2,Mao}. A number of important results have been obtained for iron-based superconductors using high-pressure techniques\cite{5_1,16,17,18,19,20,21,22}. To this end, we have performed transport and magnetic properties on NdFeAsO$_{0.9}$F$_{0.1}$ single crystal under hydrostatic pressures up to 50\,GPa. A relatively fewer studies in NdFeAsO$_{0.9}$F$_{0.1}$ have been published so far. The reason for the poor understanding of superconductivity is that all of the existing phase diagrams ($T_{c}$ vs. F-content) of the Nd-1111 systems were obtained on polycrystals\cite{Ph1,Ph2}, with all the problems involved when studying polycrystalline alloy series (foreign phases, local variation of the F-content, etc.). Very recently, we were able to synthesize high-quality F-doped crystals\cite{MA}. Additionally, the suppression of superconductivity under pressure in 1111-type of materials is not well studied. Further details on the experimental methods and a basic analysis of the sample quality can be found in Supplemental Material\cite{SI}.

In this letter, through a combined study of transport and magnetic susceptibility on NdFeAsO$_{0.9}$F$_{0.1}$, we find that while superconductivity is monotonically suppressed with increasing the pressure, the transport and optical properties reveal a prominent FL-NFL-FL crossover. A unique Pressure-temperature phase diagram of NdFeAsO$_{0.9}$F$_{0.1}$ derived from our studies shows that the NFL behavior is decoupled from superconductivity



\begin{figure}
\includegraphics[width=21pc,clip]{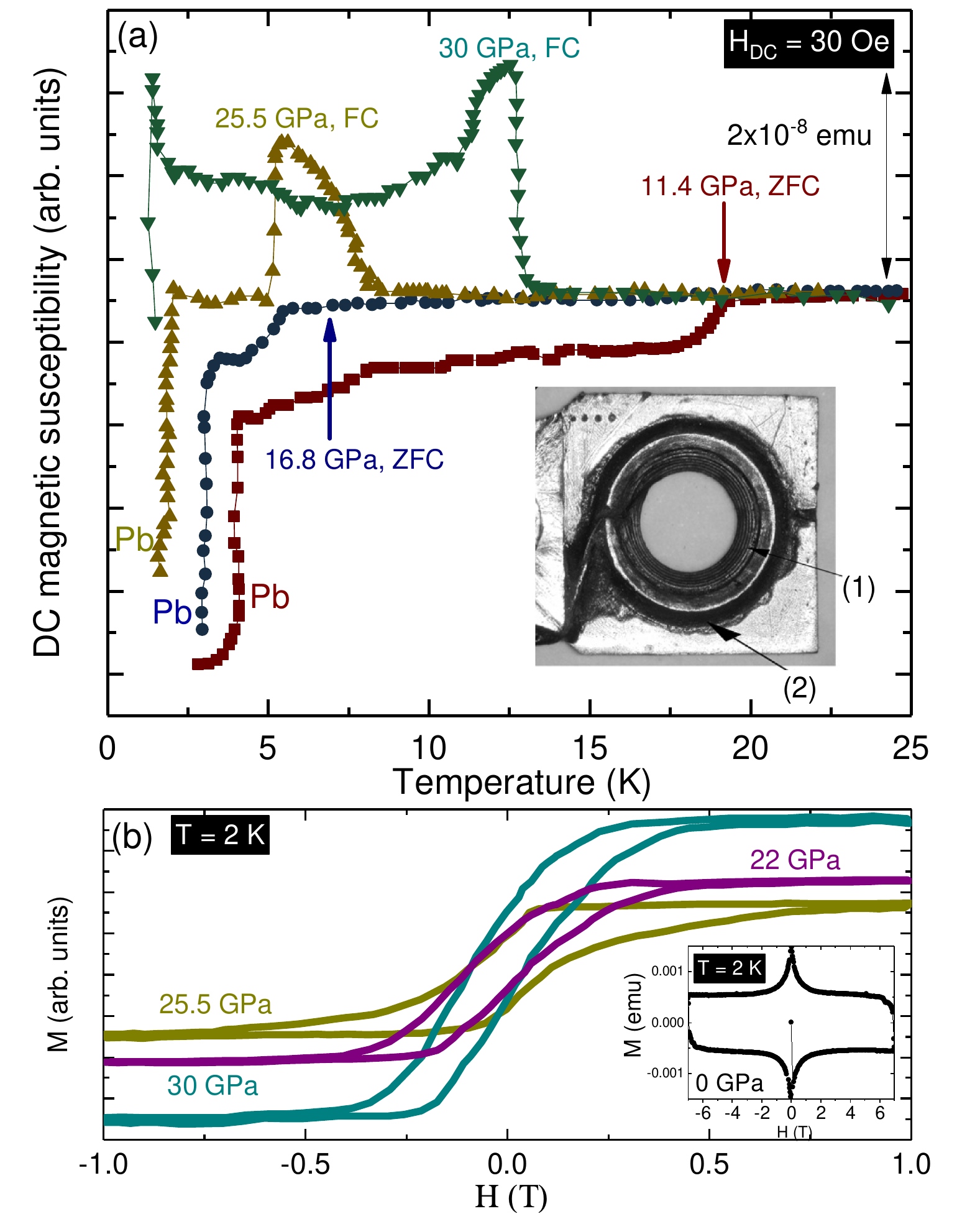}
\caption{(a) The temperature dependence of the DC-susceptibility components measured in dc field with an amplitude of 30\,Oe. The data were collected upon warming in different dc magnetic fields after cooling in a zero magnetic field. The inset illustrates a photograph of the detection coil. The inside of the main coil [20\,turns, marked as (1)] that has no bobbin was cone shaped. The compensation coil [20\,turns, marked as (2)], covered with Stycast No 2850FT, was located around the main coil, forming a concentric gradiometer~\cite{Set}. (b) The magnetic field dependence of the isothermal magnetization $M$ vs. $H$ loops measured at different pressures at 2\,K, which is consistent with a standard hysteresis loop for ferromagnets. The inset depicts the magnetic field dependence of the isothermal magnetization $M$ vs. $H$ loops measured at 2\,K up to 7\,T with the field parallel to $c$ axis.}
\end{figure}

Figure 1(a) illustrates the $T$-dependence of the in-plane resistance measurements upon heating of NdFeAsO and NdFeAsO$_{0.9}$F$_{0.1}$ single crystals. In NdFeAsO$_{0.9}$F$_{0.1}$ sample, the normal state exhibits simple metallic behavior upon cooling down from room-$T$, followed by a sharp SC transition at $T_{c} \approx$ 45.4\,K, which is in agreement with the magnetization data\cite{MA}. The $T_{c}$ is monotonically suppressed by increasing pressure up to 19.4\,GPa, which can be seen more clearly from the $\rho(T)$ data below 60\,K as shown in the inset of Fig.\,1b. Here, we define the onset $T_{c}^{onset}$ as the temperature where $\rho(T)$ starts to deviate from the extrapolated normal-state behavior, and determine $T_{c}^{zero}$ as the zero-resistivity temperature. As can be seen, upon increasing pressure to 19.4\,GPa, $T_{c}^{onset}$ is suppressed gradually to $\sim$ 5\,K and $T_{c}^{zero}$ can hardly be defined down to 2\,K, the lowest temperature in the present study. The pressure coefficient, d$T_{c}$/d$P$, is found to be around -2.2\,K/GPa. Interestingly, when increasing pressure between 22.9 and 50\,GPa, $T_{c}$ is suppressed and a broad transition appears at low temperatures as shown in Fig.\,1(c). A closer inspection of the $\rho(T)$ data in Fig.\,1(b) and Fig.\,1(c) also reveals a gradual evolution of the temperature dependence of normal-state resistivity under pressure, which will be discussed in details below.

To clarify the features of the pressure induced phase transition, the dc magnetic susceptibility was investigated under high pressure. In Fig.\,2(a), we show the temperature dependence of the DC-susceptibility components measured in dc field with an amplitude of 30\,Oe. The pressure was determined by the shift of the SC $T_{c}$ of lead located in the gasket hole. The $T$-dependence of magnetization was taken upon warming after field cooling shows a strong increase toward low temperatures upon entering the FM phase. A significant and rapid increase of the susceptibility with increasing the pressure is found. The $T_{c}$ is monotonically suppressed by increasing pressure as shown in resistivity [as discussed above]. This can be seen from  $P$ = 11.4 and 16.8\,GPa data, in which the Meissner signal was observed together with that of lead in the zero-filed cooling. Upon further increasing the pressure, at $P$ = 25.5 and 30\,GPa, the magnetic anomaly is enhanced in the field-cooled scenario upon entering the ferromagnetic phase (FM). In order to confirm this point, we plot in Fig.2(b), the magnetic field dependence of the isothermal magnetization $M$ vs. $H$ loops measured at different pressures at 2\,K. The high pressure behavior for the $M$-vs. $H$ loop reveals a standard magnetic hysteresis loop  for a FM material. While at ambient pressure (inset of Fig.\,2(b)), magnetic hysteresis loop is almost symmetric about the horizontal axis, which indicates that the hysteresis in the crystal arises mainly from bulk flux pinning rather than from the surface barrier. The saturated high-field magnetization increases with increasing pressure, signifying a stabilization of ferromagnetism under pressure.

\begin{figure}
\includegraphics[width=22pc,clip]{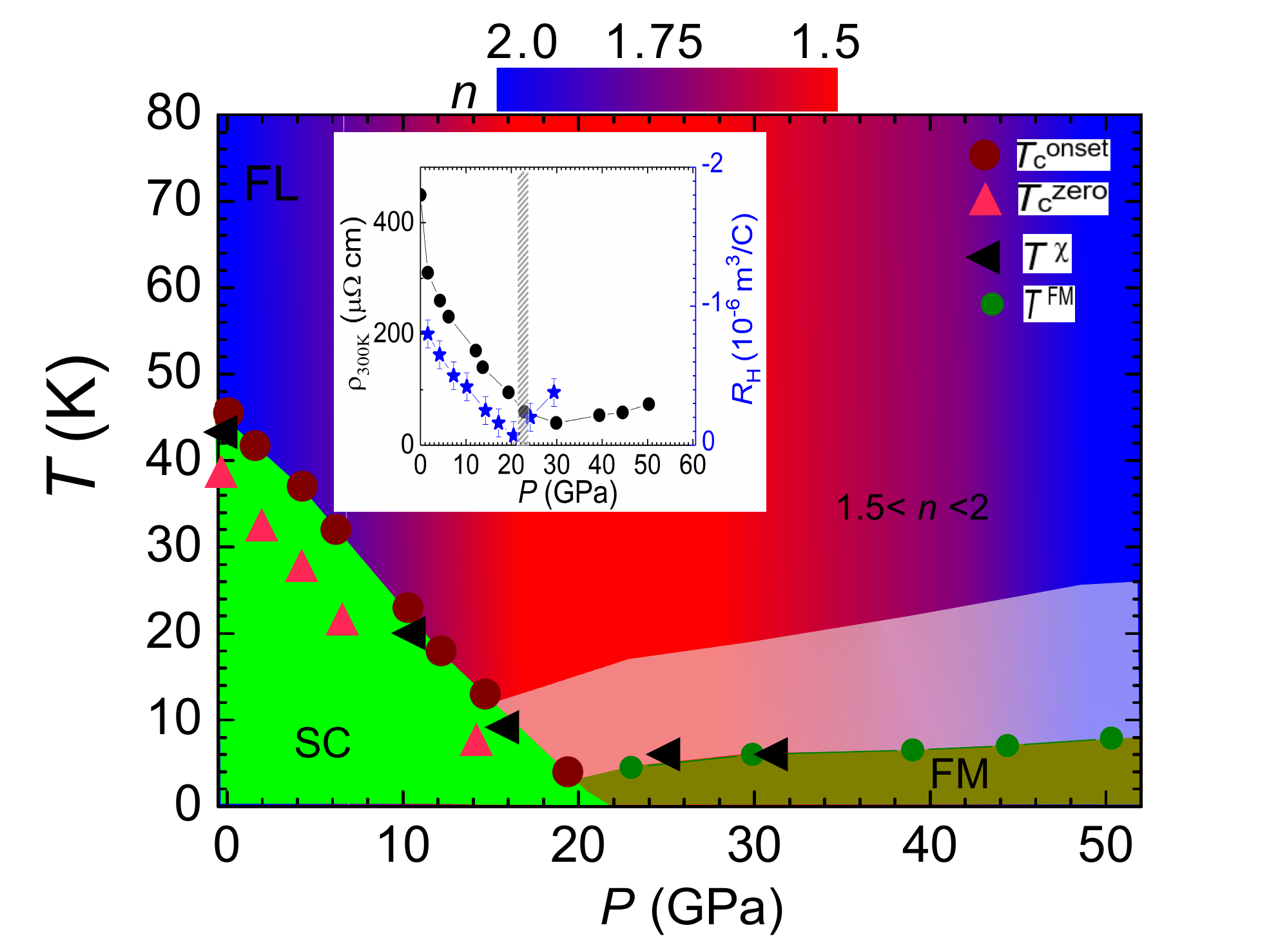}
\caption{Pressure-temperature ($P$-$T$) phase diagram of NdFeAsO$_{0.9}$F$_{0.1}$. Pressure dependence of the SC transition temperatures $T_{c}$s and a contour color plot of the normal-state resistivity exponent $n$ up to 50\,GPa. The temperature dependence of $n$ are extracted from $\rho (T)$ = $\rho_{0}$ + $AT^{n}$ for each pressure. The values of $T_{c}^{onset}$, $T_{c}^{zero}$, $T_{c}^{FM}$ and $T_{c}^{\chi}$ were determined from the high-pressure resistivity and DC magnetic susceptibility. Above $P_{c}$, local ferromagnetic order from the Nd-moments appear, observed in resistivity and susceptibility. The area above the FM regime is obtained from the blue arrows in Fig.\,4(g-h). The inset illustrates the pressure dependence of the Hall coefficient, $R_{H}$, and $\rho_{300K}$. $R_{H}$ is extracted from the transverse resistivity, $\rho _{xy}$, see Fig.\,3\cite{SI}.}
\end{figure}

\begin{figure*}[tbp]
\includegraphics[width=40pc,clip]{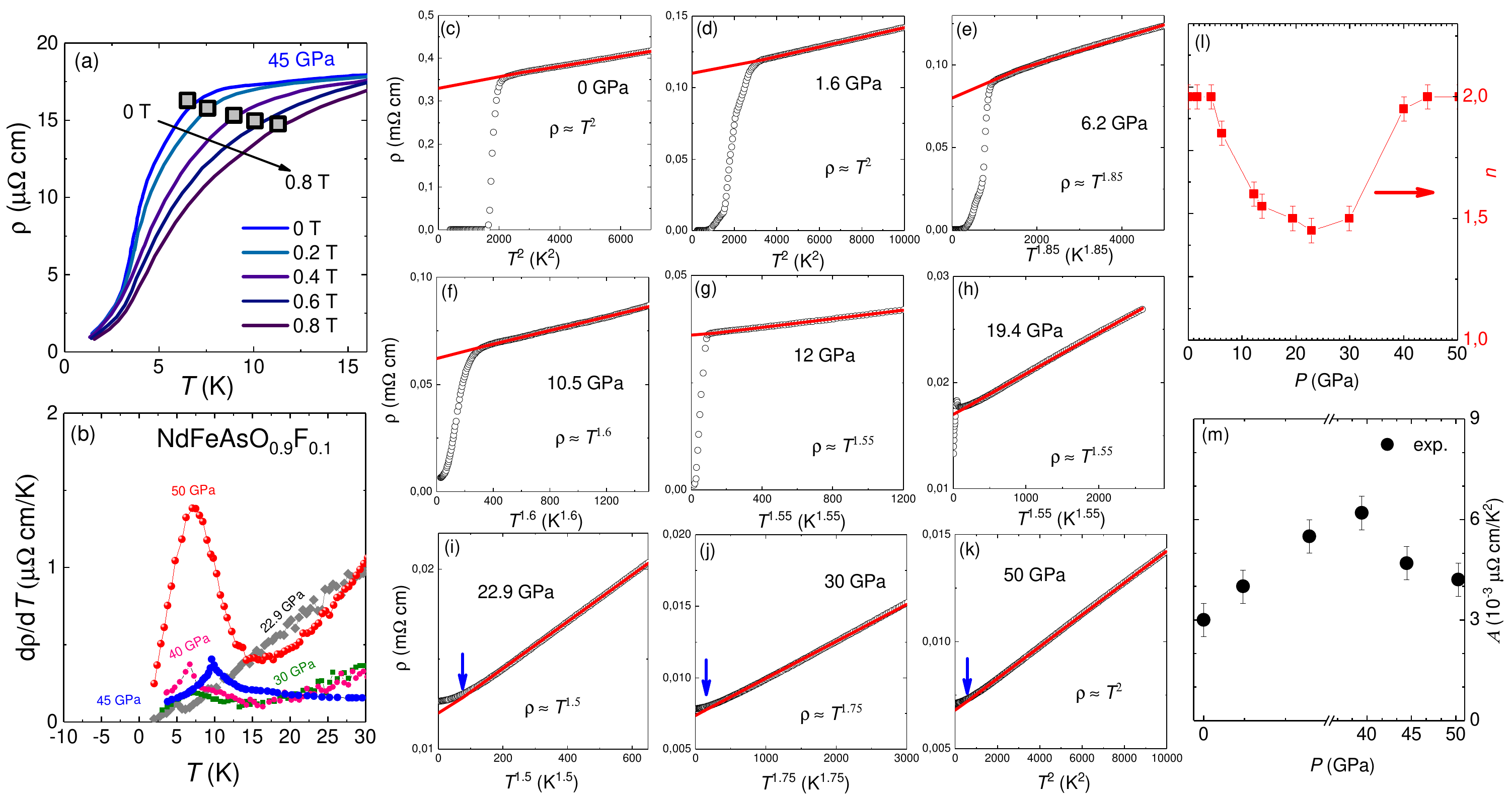}
\caption{\label{fig:wide} (a), illustrates the resistivity data at different magnetic fields under 45\,GPa. The magnetic field broadens the magnetic transition, it also slightly increases the transitions. (b), $T$-dependence of the derivative of the resistivity d$\rho$/d$T$ up to 50\,Gpa. (c-k), Resistivity as a function of $T^{n}$ for different $P$ values. $n$ is the power determined from the single power-law fit to the resistivity as a function of $T$ presented in (l). The straight solid line in each panel is linear fit. (m) The parameter $A$ is obtained by fitting the normal state resistivity data below 80\,K by using the formula: $\rho (T)$ = $\rho_{0}$ + $AT^{n}$. The blue solid line is the fitting results with the formula: $A$=[0.049-0.034 x ln(P-21)]$^{2}$.}
\end{figure*}

The pressure dependencies of the obtained $T_{c}^{onset}$, $T_{c}^{zero}$, $T_{c}^{FM}$ and $T_{c}^{\chi}$ for the studied NdFeAsO$_{0.9}$F$_{0.1}$ are summarized in Fig.\,3, which evidenced explicitly the gradual suppression of the SC phase followed by the appearance of the FM phase above $P_{c} \sim$ 21\,GPa. As can be seen, upon lowering the temperature, $\rho(T)$ displays a broad anomaly which reflects the onset of ferromagnetic ordering. The evolution under pressure of this anomaly can be clearly seen in the derivative of the high pressure data as illustrated in Fig.\,4(b). The nature of this magnetic transition is ferromagnetic, as can be seen from the $T$-dependence of the in-plane resistivity at different magnetic fields at 45\,GPa [see Fig.\,4(a)], which show that the magnetic field broadens and slightly increase the magnetic transition. A square represents the transition temperature for each field calculated from the higher temperature peak of the derivative d$\rho$/d$T$. Such an evolution of SC and FM phases is as it was observed in other Fe-based superconductors~\cite{Q5,Q6}. Since the Ruderman-Kittel-Kasuya-Yosida (RKKY) exchange coupling between the Nd local moments is oscillatory with the distance, starting with a FM coupling at low distances, the general trend towards ferromagnetism under pressure is expected\cite{Q5}. Interestingly, by using a minimal multiband model, it has recently been shown that the Fermi surface nesting has strong influence on the RKKY interaction\cite{Q5,Q5_1}. In CeFeAsO$_{1-x}$F$_{x}$ the suppression of the AFM ordering alone is not sufficient for the emergence of Ce-FM ordering as shown by F-doping studies\cite{Q5_NM}. Therefore, the origin for this behavior is more complex than a possible simple sign change of the RKKY interaction\cite{Q5_2}.

In order to gain further insights into the peculiar non-Fermi-liquid behavior, we have measured the magnetic field dependence of Hall resistivity ($\rho _{xy}(H)$) at different pressure. We noticed that all curves in Fig.\,3\cite{SI} have sub-linearity versus the magnetic field. Upon increasing pressure, all curves exhibit a negative slope in the whole investigated magnetic-field range. The inset of Fig.3 presents the pressure dependence of the Hall coefficient, defined as the field derivative of $\rho _{xy}(H)$, $R_{H} \equiv$ $\rho _{xy}(H)$/d$H$, as the slope of a linear fitting to $\rho _{xy}(H)$, see Fig.\,3\cite{SI}. In fact, the negative sign of $R_{H}$ is clearly in the whole pressure range, suggesting that the electron-type carriers dominate the charge transport in NdFeAsO$_{0.9}$F$_{0.1}$ system. As can be seen, $R_{H}$ is negative, and its magnitude first increases slightly with pressure and then experiences a quick reduction above 20\,GPa. Such a significant change in the pressure dependence of $R_{H}$ reflects a pronounced change in the band structure of NdFeAsO$_{0.9}$F$_{0.1}$ under pressure. The emergence of FM and the change
of the $R_{H}$ -pressure dependence at the same pressure indicate a close connection between them. Additionally, we noticed that both $R_{H}$ and $T_{c}$ showing similar pressure dependence behavior.

To uncover the origin of such crossover from Fermi liquid to non-Fermi liquid behavior under pressure, we have investigated the normal-state properties, which are usually correlated tightly with the SC states for unconventional superconductors. A distinct change on the temperature dependence of normal-state resistivity has already been noticed in Fig. 1b. To quantify this evolution, we fit the $\rho (T)$ to a single power law $\rho (T)$ = $\rho_{0}$ + $AT^{n}$, [see Fig.\,4(c-k)], in a sliding window width $\Delta T$=20\,K, returning the exponent $n$ and the residual resistivity $\rho_{0}$. The evolution of $n$ with doping is summarized in Fig.\,4(l). In NdFeAsO$_{0.9}$F$_{0.1}$ (from 0 to 4 GPa), the exponent $n$ is found to be 2 below 100\,K, i.e., the resistivity varies quadratically with temperature, indicating a FL normal state. Upon further increasing the pressure, $n$ decreases, reaching a minimum of 1.5 at $P$ = 14\,GPa. $n$ begins to increase and recovers to a value of 2 again at about $P$ = 40\,GPa. The minimum of $n$ near $P_{c}$ = 21\,GPa is might be related to a change of the electronic structure. These results of $n$ illustrate that a pressure induced FL-NFL-FL crossover appears at the high-pressure side of the SC dome in NdFeAsO$_{0.9}$F$_{0.1}$. This indicates that the FL-NFL-FL crossover in the investigates system is not tied to the impurity level. Simultaneously, the absolute value of the room temperature resistivity is strongly pressure dependent, see the inset of Fig.\,3.

On the other hand, in strongly correlated electron systems, the coefficient $A$ of the $T^{2}$ is often scaled as the strength of the electronic correlations. The sudden change in $A$ reflects the reconstruction of the Fermi surface topology\cite{8}. We found that the parameter $A$ decreases with increasing pressure ($P>$ 40\,GPa) as shown in Fig.4(j). If we assumed that the Kadowaki-Woods ratio~\cite{Q3_1} holds under pressure similar to the BaFe$_{2}$(As$_{1-x}$P$_{x}$)$_{2}$ system~\cite{N6}, the parameter $A$ should be proportional to the square of the effective mass $m^{*}$. Since the effective mass should vary as -$\ln(g-g_{c})$, where $g$ is tuning parameter and $g_{c}$ is the critical tuning parameter~\cite{Q4}, we would expect $A$ to be proportional to $[\alpha-\beta \ln(P-P_{c})]^{2}$. By fixing the critical pressure $P_{c}$ as 21\,GPa, we can fit the pressure dependence of parameter $A$. The blue solid curve is the fitting result of the data [see Fig.4(j)]. The fitting parameters $\alpha$ and $\beta$ are 0.049 and 0.034, respectively.




Finally, it is noteworthy that temperature dependence of resistivity in the normal-state of the overdoped La$_{2-x}$Sr$_{x}$CuO$_{4}$ shows $T^{1.6}$ at the end of the SC dome, which has been attributed to quantum criticality\cite{8}. In accordance with this picture and to our observation of similar power-law behavior near the border of SC dome in the investigated system NdFeAsO$_{0.9}$F$_{0.1}$ points to the plausible common physics that awaits for in-depth explorations in future. Experimentally, these results should stimulate new investigations on NdFeAsO$_{1-x}$F$_{x}$ and might also guide explorations to offer an important clues for discussing the unconventional origins of the high-$T_{c}$ superconductivity in this class of materials.


We acknowledge Ahuja Rajeef, Roser valenti and Ruediger Klingeler for discussions. We are grateful for support from DFG, Deutsche Forschungsgemeinschaft through MO 3014/1-1. This work has been supported by Ministry of Education and Science of Russia through NUST (MISiS) Grant No. K2-2017-084 and by the Act 211 of the Government of Russia, contract 02.A03.21.0004.

\end{document}